# Solitons in geometric potentials

Yaroslav V. Kartashov,[1] Alexander Szameit,[2] Robert Keil,[2] Victor A. Vysloukh,[1] and Lluis Torner[1]

[1]ICFO-Institut de Ciencies Fotoniques, and Universitat Politecnica de Catalunya, Mediterranean Technology Park, 08860 Castelldefels (Barcelona), Spain
[2]Institute of Applied Physics, Friedrich-Schiller-Universität Jena, Max-Wien-Platz 1, 97743 Jena
*Corresponding author: Yaroslav.Kartashov@icfo.es

We show that the geometrically-induced potential existing in undulated slab waveguides dramatically affects the properties of solitons. In particular, whereas solitons residing in the potential maxima do not feature power thresholds and are stable, their counterparts residing in the potential minima are unstable and may exhibit a power threshold for their existence. Additionally, the geometric potential is shown to supports stable multipole solitons that cannot be supported by straight waveguides. Finally, the geometric potential results in the appearance of the effective barriers that prevent transverse soliton motion.

It is conventional wisdom that a periodicity of a physical setting affecting the evolution of a wavefunction results from a change of the underlying structure of the medium, e.g. the atomic structure on the microscopic scale or the density on a macroscopic scale. However, it is often not properly appreciated that geometry may induce similar effects. For example, the curvature of a medium, embedded in a higher-dimensional space, is known to impact the wave function of a quantum particle [1,2]. Also, an electron confined to a periodically curved surface senses a periodic frictional potential which acts as a topological crystal [3]. The physics of geometric potentials is of major importance in the understanding and control of the properties of novel low-dimensional functional materials, like curved carbon nanotubes and DNA wires [4]. The impact of topological potentials is a generic wave phenomenon. In this context, optics has offered in recent years a fascinating laboratory tool to investigate classical analogues of otherwise inaccessible quantum-mechanical and relativistic effects (see, e.g., [5]). Recently for the first time such a geometric potential was experimentally demonstrated in optics for a curved two-dimensional plane which is embedded in the three-dimensional surrounding space [6,7]. With these results, a so-called *photonic topological crystal* could be realized. As a model system for the topological potential, an undulated slab waveguide was used, that resembled an accepted procedure to confine a quantum particle on a surface that was originally proposed by Jensen et al [1,2]. However, in [7] only linear propagation effects were considered, such as topological Bloch oscillations and Zener tunnelling. The interplay between self-action effects and potentials of geometric origin was never addressed before.

In this Letter we analyze theoretically nonlinear light evolution in undulated slab waveguides, and investigate the impact of geometric potential arising due to waveguide curvature on the evolution of optical solitons. In particular, we show that such potentials dramatically affect power thresholds and stability of solitons residing in sections with different waveguide curvature.

We describe the propagation of laser radiation along the $\xi$ axis of focusing cubic medium with an imprinted transverse modulation of refractive index by the Schrödinger equation for the dimensionless field amplitude $q$:

$$i\frac{\partial q}{\partial \xi} = -\frac{1}{2}\left(\frac{\partial^2 q}{\partial \eta^2} + \frac{\partial^2 q}{\partial \zeta^2}\right) - pR(\eta,\zeta)q - q|q|^2, \quad (1)$$

where the propagation distance $\xi$ is scaled to the diffraction length; the transverse coordinates $\eta,\zeta$ are scaled to the characteristic width; $p$ is the lattice depth; the function $R(\eta,\zeta)$ describes the transverse refractive index variation. Here we consider a quasi-one-dimensional slab waveguide whose center experiences periodic oscillations in the transverse plane according to the law $\zeta = a\sin(\Omega\eta)$, where the parameters $a$ and $\Omega$ stand for the bending amplitude and frequency, respectively [see Fig. 1(a) with an example of such a waveguide]. The geometric potential is thereby inversely proportional to the local bending radius; hence it is maximal where the bending is largest and vanishes in the straight section of the slab waveguide. Since we aim to realize a potential of purely geometrical origin the width of quasi-one-dimensional waveguide is supposed to be constant in the direction normal to the instantaneous tangential line to the curve where the waveguide center is located. Thus, in the curvilinear coordinate system $(u,v)$, where $u$ is the coordinate along the curved array axis and $v$ is the coordinate in the orthogonal direction, the waveguide shape is given by $R = \exp(-v^4/d^4)$, where $d$ is the waveguide width. Due to constant width of the waveguide in normal direction the total potential affecting the propagation of laser radiation does not contain a contribution stemming from local variation of effective refractive index, but it does contain purely geometrical contribution stemming from instantaneous waveguide curvature. Notice that curved waveguides of constant width can be fabricated with the aid of femtosecond laser-writing technique [8]. Further we set $\Omega = 1$ (which corresponds to bending period $\sim 31.4$ $\mu$m ), $d = 0.4$ (the waveguide with the 4 $\mu$m width), $p = 8$ (that is equivalent to refractive index contrast $\delta n \sim 10^{-3}$ at wavelength $\lambda = 800$ nm ), and we investigate the impact of bending amplitude $a$ that strongly affects the strength of the geometric potential on the properties of solitons supported by curved waveguide.

The solitons of Eq. (1) can be found in the form $q(\eta,\zeta,\xi) = w(\eta,\zeta)\exp(ib\xi)$, where the function $w(\eta,\zeta)$ describes soliton shape and $b$ is the propagation constant. In order to analyze stability of such states we write perturbed field as $q = [w + u\exp(\delta\xi) + iv\exp(\delta\xi)]\exp(ib\xi)$ with $u, v \ll w$ being real and imaginary parts of perturbation, respectively, and $\delta$ being the perturbation growth rate. The substitution of such field into Eq. (1) and linearization yield a linear eigenvalue problem:

$$\delta u = -\frac{1}{2}\left(\frac{\partial^2 v}{\partial \eta^2} + \frac{\partial^2 v}{\partial \zeta^2}\right) + bv - pRv - w^2 v,$$
$$\delta v = +\frac{1}{2}\left(\frac{\partial^2 u}{\partial \eta^2} + \frac{\partial^2 u}{\partial \zeta^2}\right) - bu + pRu + 3w^2 u,$$
(2)

that we solved numerically to find growth rates $\delta$.

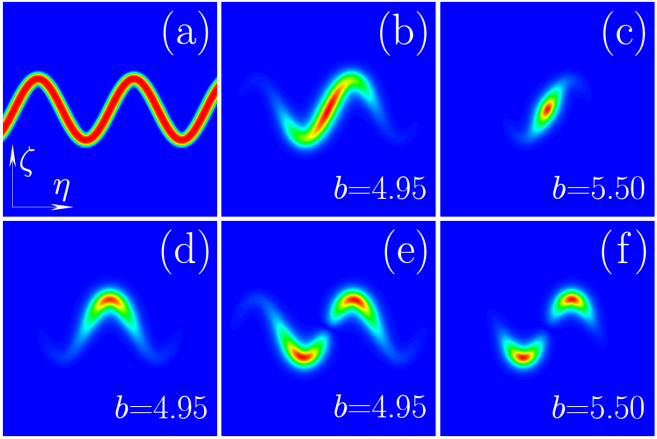

Figure 1. (a) Example of periodically curved quasi-one-dimensional channel. Profiles of fundamental solitons centered on the points with zero (b),(c) and maximal (d) channel curvature. Profiles of dipole solitons supported by curved channel are shown in (e) and (f). In all cases field modulus distributions are shown, while $a = 2$.

One of the central results of this Letter is that a geometric potential stemming from local waveguide curvature dramatically affects the possible locations, shapes, and properties of solitons supported by bent waveguide. There are several locations where simplest fundamental solitons can reside that include the points with vanishing waveguide curvature at $\eta = 2n\pi/\Omega$ [see Figs. 1(b) and 1(c) for corresponding soliton shapes], and the points where the waveguide curvature is maximal at $\eta = (2n+1)\pi/\Omega$, where $n$ is an integer number [see Fig. 1(d)]. Such states always exist above certain cutoff propagation constant $b_{co}$. The behavior of soliton solution at $b \to b_{co}$ depends dramatically on the amplitude of waveguide bending (i.e. on the strength of geometric potential). When $a$ is relatively small both solitons residing around points with zero and maximal curvature expand drastically along the waveguide when $b \to b_{co}$ [see Figs. 1(b) and 1(d)]. When $a \sim 4$, solitons residing at points with maximal curvature are always well localized, whereas solitons residing in zero curvature point show a tendency for splitting into two bright spots shifted toward maximal curvature points. Far from the cutoff value of $b$ such fundamental solitons transform into well-localized bright spots [Fig. 1(c)]. Soliton solutions can be characterized by the dependence of energy flow $U = \int\int_{-\infty}^{\infty}|q|^2\, d\eta d\zeta$ on propagation constant. While at low values of $a$ solitons at zero curvature points do not feature any energy flow threshold (i.e. $U \to 0$ when $b \to b_{co}$), at moderate and high $a$ values when geometric potential becomes sufficiently strong, such solitons require a certain minimal energy flow for their existence as shown in Fig. 2(a). At the same time, fundamental solitons residing in the points with maximal waveguide curvature never feature energy flow threshold for any $a$, as shown in Fig. 2(b), curve 1 (by analogy with on-site discrete solitons [9]). This is a direct indication of the presence of the geometric potential that makes the properties of solitons residing in various points in the waveguide remarkably different – something that does not occur in straight waveguide where soliton properties do not depend on the soliton position. Besides the difference in thresholds, we found that while solitons residing in the points with maximal curvature are always stable, their counterparts on zero curvature points are unstable [see Fig. 2(c) for the dependence of perturbation growth rate on $b$ for such states]. This instability causes a soliton drift and it is not captured by the standard Vakhitov-Kolokolov stability criterion.

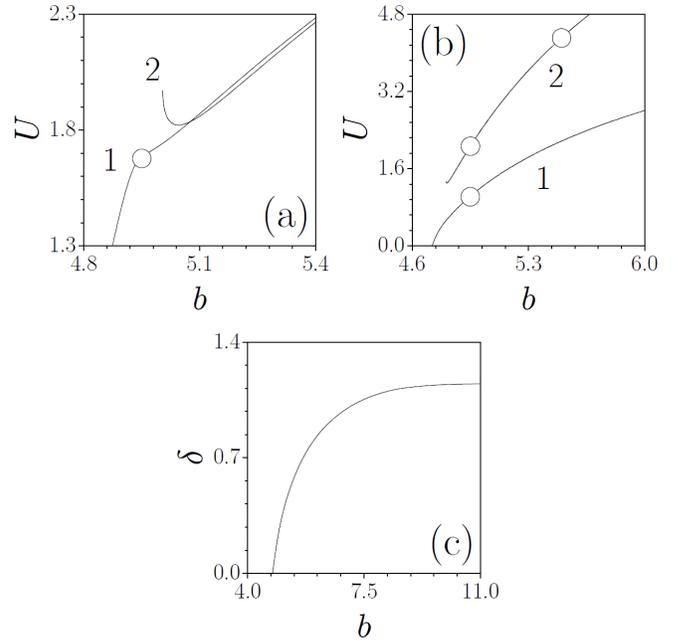

Figure 2. (a) $U$ versus $b$ for fundamental soliton residing in the point with zero curvature at $a = 2$ (curve 1) and $a = 3$ (curve 2). Circle corresponds to soliton in Fig. 1(b). (b) $U$ versus $b$ for fundamental soliton residing in the point with maximal curvature (curve 1) and for dipole soliton (curve 2) at $a = 2$. Circles correspond to solitons from Figs. 1(d)-1(f). (c) $\delta$ versus $b$ for fundamental soliton in zero curvature point at $a = 2$.

Due to the geometric potential that is capable of compensating repulsive forces between out-of-phase spots the curved waveguide can support multipole solitons that are impossible in straight waveguide [see Figs. 1(e) and 1(f) showing simplest dipole soliton]. The soliton poles that are out-of-phase are located in the maximal curvature points. Multipole solitons always feature energy flow threshold even if corresponding fundamental states are thresholdless [Fig. 2(b), curve 2]. The localization of spots in multipole solitons grows with increase of bending amplitude $a$ or propagation constant $b$. At small values of $a \sim 1$ multipole solitons can be unstable around the cutoff on propagation constant, but they become stable with increase of $b$. For large bending amplitudes multipole solitons are stable almost in the entire existence domain. Figure 3 illustrates the instability of perturbed fundamental solitons residing in zero-curvature points [this instability causes a shift of solitons to the point where the curvature is maximal, as shown in Fig. 3(a)], and stable propagation of fundamental solitons in the maximal curvature point [Fig. 3(b)] as well as dipole solitons [Fig. 3(c)].

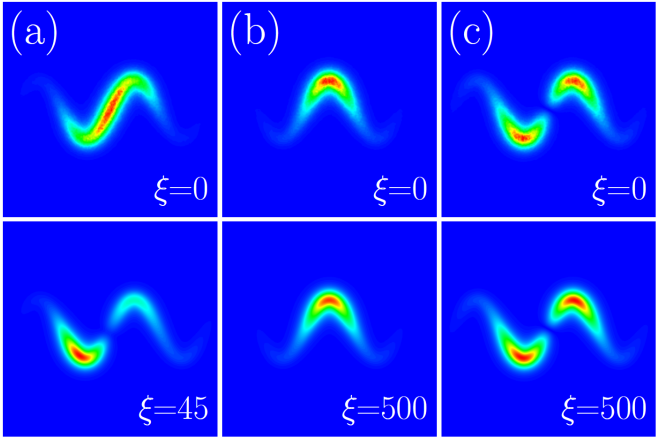

Figure 3. Decay of unstable fundamental soliton in zero-curvature point (a), and stable propagation of fundamental soliton residing in the point of maximal curvature (b) and dipole soliton (c). White noise was added into input field distributions. In all cases $b = 4.95$, $a = 2$.

Another important manifestation of a geometric potential is that it prevents the motion of solitons along the curved waveguide, in contrast to the case of straight waveguide where solitons move freely. In order to study the dependence of the effective potential barrier (also known as Peierls-Nabarro barrier [10,11]) arising due to geometric effects on bending amplitude we imposed the initial phase gradient $\exp(i\theta\eta)$ on stationary soliton residing in the maximal curvature point and determined the critical value $\theta = \theta_{cr}$ at which soliton starts moving along the waveguide [see Fig. 4(c) with an example of such a motion] as a function of the bending amplitude. For a fixed energy flow of the soliton the dependence $\theta_{cr}(a)$ is almost linear as shown in Fig. 4(a). Surprisingly, a critical phase tilt $\theta_{cr}$ required for setting the soliton into motion diminishes with increase of its energy flow [Fig. 4(b)]. This is in drastic contrast to the case of periodic waveguiding systems where the height of the effective potential barrier preventing solitons from motion in the transverse plane usually rapidly grows with $U$ [10,11]. One can suppose that $\theta_{cr} \to 0$ when the energy flow of the solitons approaches the Townes soliton energy $U_T = 5.85$ at $b \to \infty$. In this limit the nonlinear contribution to the refractive index highly exceeds the linear refractive index modulation and the solitons becomes so narrow that it does not feel the local curvature of the undulated waveguide anymore.

Summarizing, we studied the existence and properties of solitons propagating in undulated waveguides that induce a purely geometric potential. We showed that the stationary and dynamical properties of solitons existing in such structures depend dramatically on the amplitude of waveguide bending. Our findings stress the importance of geometrically-induced potentials to control localized light and matter states.

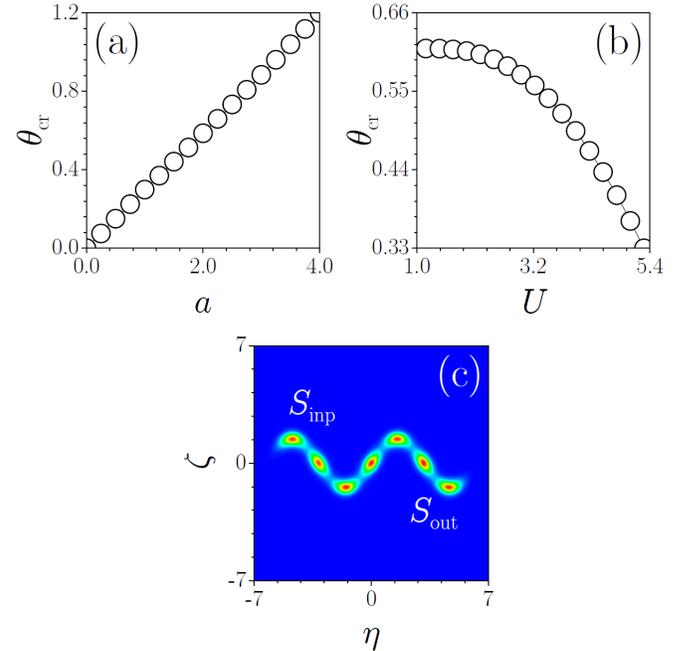

Figure 4. Critical input tilt (a) versus bending amplitude for solitons with $U = 2.8$ and (b) versus soliton's energy flow at $a = 2$. (c) Snapshot images showing motion of soliton with $U = 2.8$ along curved channel with $a = 1.5$ for $\theta > \theta_{cr}$. Labels "$S_{inp}$" and "$S_{out}$" denote input and output soliton positions, respectively.